# AI-Assisted Writing in Education: Ecosystem Risks and Mitigations

Antonette Shibani
Transdisciplinary School
University of Technology Sydney
Australia
antonette.shibani@uts.edu.au

Simon Buckingham Shum
Connected Intelligence Centre
University of Technology Sydney
Australia
simon.buckinghamshum@uts.edu.au

## ABSTRACT

While the excitement around the capabilities of technological advancements is giving rise to new AI-based writing assistants, the overarching ecosystem plays a crucial role in how they are adopted in educational practice. In this paper, we point to key ecological aspects for consideration. We draw insights from extensive research integrated with practice on a writing feedback tool over 9 years at a university, and we highlight potential risks when these are overlooked. It informs the design of educational writing support tools to be better aligned within broader contexts to balance innovation with practical impact.

## CCS CONCEPTS

•Applied computing ~ Education ~ Interactive learning environments •Social and professional topics ~ Computing / technology policy

## KEYWORDS

Artificial Intelligence, Writing, Assistant, Feedback, Ecosystem, Risks, Mitigations, Tools, Education, Classroom, Adoption, LLM

## 1 Introduction

Software has gradually become an integral part of the writing process [1], such that authors are now enmeshed with technology in "socio-cyborgian activity systems" [2], foregrounding the need to teach the next generation to critically engage with Artificial Intelligence feedback [3]. Advances in generative AI (GenAI) are propelling R&D into intelligent and interactive writing assistants (IIWAs) enabling innovative forms of writing feedback [4, 5] and human-AI co-writing [6]. However, as startups launch new IIWA apps almost every month, we must avoid the potential pitfalls of *"technosolutionism"*, the entrepreneurial mindset that complex issues can be solved by technology alone [7]. Rather, we must learn from the long history of prior IIWA research [1, 8] and "Macro-HCI" approaches [9], to consider the broader socio-technical *ecosystem* in which the tool must become embedded, where multiple actors define the context in which the IIWA is used [5]. In this position paper, we draw on our experiences refining an educational writing tool, to illustrate how designing this ecosystem has proven critical to transitioning it from research prototype to product, deployed at scale in authentic classroom contexts [10, 11]. Conversely, failure to do so hinders adoption.

## 2 Ecosystem dimensions of AI Writing Assistants

In a recent review that mapped the design space of IIWAs [5], the ecosystem dimensions were identified as important, and involve: *"the overarching sociotechnical context in which the writer and the tool are situated, encompassing a range of complex, interdependent actors that frequently play a role in the functioning of the writing assistant"*. However, there was sparse empirical evidence in the literature on how this ecosystem manifests, given the novelty of GenAI IIWAs in particular.

In higher education, scaling up AI research prototypes faces 'growing pains' [10] in authentic contexts. To help IIWA designers avoid repeating this history, we focus attention on the ecosystem. Since 2015, we have been evolving an educational IIWA to the status of an internal 'product' in the university[1], working closely with educators and students to co-design the tool, integrating it into the institution's technical and information ecosystem. This tool, *AcaWriter* provides automated feedback on students' academic and reflective writing by highlighting salient rhetorical structures using Natural Language Processing (NLP) rules [12]. It has been implemented for several disciplines and genres at university, with contextualized feedback aligned to Law, Accounting, Pharmacy, and Engineering professional contexts, tested out in authentic classrooms with educator input [12, 13]. Considering the wider ecosystem factors has led to institutional adoption and learner uptake for impactful outcomes [14].

In response to the workshop call to consider the 'dark sides' of IIWA, Table 1 summarises key risks if ecosystem dimensions are ignored or poorly executed, and examples of mitigating measures taken in our longitudinal project. While recent advances in Large Language Models (LLMs) offering more powerful capabilities raise new challenges (such as how learners might need additional support to learn how to critically engage with it [15]), we believe that lessons from past writing assistants such as AcaWriter can inform future IIWAs. Note that other important aspects of our work concerning data ethics [16], algorithmic bias, and explainability are not covered to keep a tight focus on the ecosystem dimensions, which often otherwise remain hidden.

---

[1] https://acawriter.uts.edu.au/

Table 1: *Ecosystem* dimensions from the IIWA design space [5]: Sample risks and mitigations in higher education

| Dimension | Options | Potential ecosystem risks and the mitigating approach taken in the AcaWriter project |
|---|---|---|
| **Digital Infrastructure**<br><br>*What compatibility issues are considered?* | • Usability consistency<br>• Technical interoperability | **Risks:** IIWA is siloed from other technical services; not integrated into the wider information support system for writing; user interface is inconsistent with current tools, or violates usability conventions<br><br>**Mitigation:**<br>• Student signs in with standard university authentication<br>• Bi-directional website crosslinks between the app and human support services for writing<br>• Graphical user interface consistent with word processing conventions |
| **Access Model**<br><br>*How does the access model influence design decisions, or openness of data/algorithms?* | • Free and open-source software<br>• Commercial software | **Risks:** Closed source black box IIWA restricts sharing of technological advancements, limiting reproducibility, transparency, and aggravating equitable access issues<br><br>**Mitigation:**<br>• Research transparency in how the app's functionality is grounded in writing theory and empirical evidence [12, 17]<br>• Publish open source code with installation support [18], and open access datasets [19]<br>• Publish open access educator resources [20] |
| **Social Factors**<br><br>*Who affects the design and use of writing assistants?* | • Design with stakeholders<br>• Design for social writing | **Risks:** A technocentric focus on new "cool" features eclipses meaningful engagement from the early design stages with stakeholders (e.g., authors, educators, coaches). IIWA fails to gain their trust and sustained adoption, e.g., incompatibilities with user needs, terminology, alignment with educational values, or poor support for social practices that emerge around the app<br><br>**Mitigation:**<br>• Use participatory design to co-design IIWA feedback, user experience, and other functionality with educators across diverse disciplines [13, 21]<br>• Pilot with students in authentic contexts over multiple years [12, 21] and evidence impact [14] |
| **Locale**<br><br>*Does the writing assistant's design take into account features of a physical locale?* | • Local writing<br>• Remote writing | **Risks:** IIWA disregards the different affordances of local versus remote writing, and has no consideration of contexts it will be used<br><br>**Mitigation:**<br>• Design specifically to support students 24/7 with automated feedback on writing, but the app does assume use of a desktop/laptop in its interface design (adaptive layout for mobile devices may be possible in future versions)<br>• Within these constraints, the app supports remote or mobile writing away from class, but its use is contextualised as far as possible in curriculum-embedded writing tasks [13]<br>• Students can save as many texts as they want, assisting in revising from any local/remote context |
| **Norms & Rules**<br><br>*What norms and rules affect the design and use of a writing assistant?* | • Laws<br>• Conventions | **Risks:** IIWA design takes no account of legal or cultural norms that impinge on how the app will be used; fails to ground technological features in theory and evidenced practice<br><br>**Mitigation:**<br>• Ground the IIWA in relevant theory of writing and pedagogy that makes sense to the teaching teams<br>• Understand and align the IIWA with disciplinary and genre conventions [13]<br>• Respect and partner with university stakeholders, not only students and educators, but senior leaders and technology gatekeepers [11] |
| **Change Over Time**<br><br>*What are the key temporal considerations when designing a writing assistant?* | • Authors<br>• Readers<br>• Writing<br>• Information environment<br>• Technologies<br>• Regulation | **Risks:** IIWA design takes no account of ecosystem dynamics, as different elements change at different rates. Quality and robustness are defined solely in terms of software, or model performance, rather than effective usage under authentic conditions. There is little or no capability to track pedagogically meaningful usage patterns, how use varies over time or demographic, and the impact on writing process and product<br><br>**Mitigation:**<br>• IIWA is evaluated in authentic contexts and changes in writing behaviour are tracked over time [12]<br>• The app is insulated from external changes (e.g., language model upgrades) to avoid unexpected changes in behaviour<br>• The app is grounded in open-source language technology that has been cleared of intellectual property concerns<br>• Analytics can be derived from fine-grained writing logs [22] and integrated with authors' self-reports and academic data [12] |

## 3 Conclusion

Building on the *ecosystem* dimensions identified in the recent interactive writing assistants design space [5], we have mapped associated risks and mitigations, drawing on our insights developing an AI writing tool and implementing it at scale in higher education over a sustained period of 9 years. The factors we discussed are particularly important for designers of *educational IIWAs* to consider, ensuring the primary goal of improving learner experience through the data and tools we develop is attained [23]. This approach shifts the focus from algorithmic perfection to aiding learning [24], and from cool features to meaningful long-term engagement as learners start using a new generation of IIWAs. We hope that these offer insights for emerging generative AI writing tools aiming to impact learners in authentic contexts. We invite future IIWA designers to consider how ecosystem dimensions play a part in their own contexts and add to the repertoire of mitigation strategies that may extend beyond educational contexts.